\documentclass{article}
\usepackage{authblk}
\usepackage{amsfonts}
\usepackage{amstext}
\usepackage{color}
\usepackage{lineno}
\usepackage{amssymb}
\usepackage{amsthm}
\theoremstyle{plain}
\definecolor{Red}{cmyk}{0,1,1,0}

\definecolor{Blue}{cmyk}{1,1,0,0}

\newcommand{\eop}{\hfill \rule{0.7ex}{1.6ex}}
\newcommand{\F}{ {\cal F} }

\newcommand{\tra}{\widehat}

\newcommand{\dsum}{\displaystyle \sum}

\newcommand{\R}{\mathbb{R}}

\newcommand{\B}{{\cal B}}

\newcommand{\N}{\mathbb{N}}

\newtheorem{theorem}{Theorem}[section]
\newtheorem{lema}{Lemma}[section]

\title{
	{\large{ \bf{A RENORMALIZATION GROUP APPROACH TO HIGHER ORDER CORRECTIONS TO THE DECAY OF
SOLUTIONS TO NONLINEAR INTEGRAL EQUATIONS}}}}

\vspace{2cm}
\date{}
\author[1]{\scshape Gast\~ao A. Braga}
\author[1]{\scshape Jussara M. Moreira} 
\author[3]{\scshape Ant\^onio Marcos da Silva}
\author[4]{\scshape Camila F. Souza}
\affil[1]{Departamento de Matem\'atica, Universidade Federal de Minas Gerais}
\affil[3]{Departamento de Matem\'atica, Universidade Federal de Ouro Preto}
\affil[4]{Departamento de Matem\'atica, Centro Federal de Educa\c c\~ao Tecnol\'ogica de Minas Gerais}

\date{}

\begin{document}
	\maketitle
	
%
%
%
%
%
%
	
	\medskip

	\def\l{\lambda}
	
	\baselineskip = 22pt
	
	\maketitle
	
	\begin{abstract}
	In this paper we employ the Renormalization Group (RG) method to study higher order corrections to the long-time asymptotics of a class of nonlinear
integral equations with a generalized heat kernel and with time-dependent coefficients. This is a follow up of the papers \cite{bib:souz-brag-more-art,bib:souz-brag-more-marginal,bib:brag-more-silv}.
	\end{abstract}
	\clearpage

\section{\large{Introduction}}
	\label{sec:intr}

In \cite{bib:souz-brag-more-art,bib:souz-brag-more-marginal}, the long-time asymptotics of solutions to the following class of nonlinear integral equations
\begin{equation}
	\label{eq:nao-lin}
	u(x,t)=\int {G(x-y, s(t))f(y)dy} +
\lambda\int_1^{t}\int {G(x-y,\phi(\tau))F(u, u_x)(y,\tau)\, dy d\tau}
\end{equation}
was investigated using the Renormalization Group (RG)
method developed by Bricmont et al.~in \cite{bib:bric-kupa-lin}.
In the above equation,  $G(x, t)$ is required to  satisfy
some properties (see conditions $\textbf{(G)}$ listed below), $s(t)$ is given by
\begin{equation}
\label{def:s(t):int}
s(t)=\int_{1}^{t}c(\tau)d\tau,
\end{equation}
where $c(t)$ is a positive function in $L^1_{loc}((1,\infty))$ and $c(t) = t^p+o(t^p)$ as $t \to \infty$, with $p>0$, $\phi(\tau)=s(t)-s(\tau)$, and $f$ is in an appropriate Banach space. In \cite{bib:souz-brag-more-art,bib:souz-brag-more-marginal}, the nonlinear term $F$ depends only on $u$ and is given by $F(u) = -\mu u^{\alpha_c} + \sum_{j >\alpha_c}{a_ju^j}$, where
 $\alpha_c=(p+1+d)/(p+1)$,
with $p$ and $d$ being positive exponents related to $s(t)$ and $G(x,t)$, respectively.
The asymptotic results in \cite{bib:souz-brag-more-art}, for $\mu = 0$ in the above definition of $F(u)$, and in  \cite{bib:souz-brag-more-marginal} for $\mu > 0$,
are described by
\begin{equation}
\label{def:marg-irre-beha}
u(x,t) \sim \frac{A}{[t\,\, r(t)]^{(p+1)/d}}G\left(\frac{x}{t^{(p+1)/d}},
\frac{1}{p+1}\right)\mbox{  when } t\rightarrow \infty.
\end{equation}
where $r(t) =1$ if $\mu=0$ and $r(t) = \ln t$ if $\mu>0$. The behavior
(\ref{def:marg-irre-beha}) is an outcome of the RG operator flow analysis around
its linearization's fixed point $G\left(x, 1/(p+1)\right)$.

The aim of this paper is to keep exploiting the RG method
to study the asymptotic behavior of solutions to (\ref{eq:nao-lin}), with the nonlinearity
 \begin{equation}
\label{def:irre-pert}
F(u, u_x)=  \sum_{j \geq\alpha}{a_ju^ju_x}
\end{equation}
analytic at the origin, with radius of convergence $r > 0$, where $\alpha$ is a positive integer greater than $\alpha_c$ and
 \begin{equation}
		\label{def:alpha-c}
		\alpha_c = \frac{d-(p+1)}{2(p+1)},
		\end{equation}
but considering the initial data $f$ under the {\em zero mass condition} $\int_\R f(x) dx = 0$.
	
Under the zero mass condition, $A=0$ in (\ref{def:marg-irre-beha}) and, in order to get the
right asymptotic behavior, it is necessary to find the next leading order contribution to the
long time behavior and this will be done by using the RG multiscale analysis.
Following \cite{bib:brag-more-silv} we show that, also in this case, there is
a well defined RG operator so that
the long time behavior of $u(x,t)$ is dictated by a fixed point of its linearization and is given by
\begin{equation}
\label{def:zero-mass-irre-beha}
u(x,t) \sim \frac{A}{t^{2(p+1)/d}}G'\left(\frac{x}{t^{(p+1)/d}},
\frac{1}{p+1}\right)\mbox{  when } t\rightarrow \infty,
\end{equation}
where $G'(x,t)\equiv \partial_x G(x,t)$ and $G'(x,1/(p+1))$ is a fixed point
of the linear RG operator.

The RG method is a
multiscale method that has shown to be quite appropriate for getting fine information
on the asymptotics of solutions to time evolution problems
Concerning the asymptotics, via the RG method, under the zero mass condition,
we point out the work \cite{bib:bona1} of Bona et al. Using the standard RG operator,
see \cite{bib:bric-kupa-lin}, Bona et al. analyzed the asymptotic
behavior of solutions to a generalized Kortweg-de Vries equation but the long time profile function
is not a fixed point of their RG operator.  Subsequently, based on the numerical results presented in
\cite{bib:brag-Furt-isa-Lee}, Braga et al. \cite{bib:brag-more-silv} redefined the RG operator to study
a generalized Burgers equation considering the zero mass.

The  kernel $G=G(x, t)$ satisfies the following conditions which we denote
by $\textbf{(G)}$:
	\begin{enumerate}
	\item[$(G_1)$]
				There are integers $q >1$ and $M>0$ such that $G(\cdot,1)\in C^{q+2}(\R)$ and
			$$
			\sup_{x\in\R}\{(1+|x|+x^2)(1+|x|)^{M+1}|G^{(j)}(x, 1)|\}<\infty, ~~~~j=0,1,..., q +2,
			$$
			where $G^{(j)}(x, 1)$ denotes the j-th derivative $(\partial_x^jG)(x, 1)$.
	\item[$(G_2)$]
	There is a positive constant $d$ such that
	$$
	G(x,t)=t^{-\frac{1}{d}}G\left(t^{-\frac{1}{d}}x,1\right),~~~~ x\in\R,~t>0;
	$$
	\item [$(G_3)$]
	$
	G(x,t)= \int_\R{G(x-y, t-\tau)G(y,\tau)dy}~~~ \mbox{ for } x\in\R \mbox{ and } t>\tau>0.
	$
\end{enumerate}

From these properties, we can now present the heuristics for obtaining the asymptotics in the linear case. Consider the solution to (\ref{eq:nao-lin}) with $\lambda=0$ and $s(t) \approx \frac{t^{p+1}}{p+1}$ for large $t$. Using condition $(G_2)$ in Fourier space $\widehat{G}\left(\omega,\frac{t^{p+1}}{p+1}\right)=\widehat{G}\left(t^{\frac{(p+1)}{d}}\omega,\frac{1}{p+1}\right)$, we can write
\begin{eqnarray*}
u(x,t) =
\frac{1}{2 \pi}
\int_{|\omega|<\epsilon} \widehat{G}\left(t^{\frac{(p+1)}{d}}\omega,\frac{1}{p+1}\right) e^{i\omega x}\hat{f}(\omega)d\omega \\ +
\frac{1}{2 \pi}
\int_{|\omega|>\epsilon} \widehat{G}\left(t^{\frac{(p+1)}{d}}\omega,\frac{1}{p+1}\right) e^{i\omega x}\hat{f}(\omega)d\omega .
\end{eqnarray*}
The smoothness of $G(x,t)$ implies a certain polynomial decay rate of $\tra G(\omega,1)$
	so that, after expanding $\hat{f}$ around $\omega = 0$, we get
	$$
	u(x,t) \approx \frac{1}{2 \pi}
	\int_{|\omega|<\epsilon}\widehat{G}\left(t^{\frac{(p+1)}{d}}\omega,\frac{1}{p+1}\right) e^{i\omega x}
	\left[\hat{f}(0)+\omega \hat{f}'(0) +
	O\left( \omega^2
	\right)\right]
	d\omega.
	$$
Assuming the zero mass hypothesis and dropping off $O\left( \omega^2
\right)$ terms from the above integral, after a change of variables,
$$
u(x,t) \approx \frac{1}{2 \pi} \left[\frac{1}{t^{(p+1)/d}}\right]^2
\int_{\left|\frac{\omega}{t^{(p+1)/d}}\right|<\epsilon}
\omega \,\, \tra G\left(\omega,\frac{1}{p+1}\right)\,\, e^{i\left(\frac{\omega}{t^{(p+1)/d}}\right) x}\,\,  \hat{f}'(0)\,\, d\omega,
$$
which gives us the right asymptotic behavior of $u$
\begin{equation}
	\label{aprox-formal}
	u(x,t) \approx
	\frac{A}{t^{2(p+1)/d}}G'\left(\frac{x}{t^{(p+1)/d}}\right),
	\ t \gg 1,
\end{equation}
where $A = -i\hat{f}'(0)$
and $G'\left(x \right)$ stands
for the $x$-derivative of $G\left(x, \frac{1}{p+1}\right)$. The conclusion, from
(\ref{aprox-formal}), is that as $t\to\infty$, the solution to the linear problem ((\ref{eq:nao-lin}) with $\lambda=0$)
decays with rate $t^{2(p+1)/d}$, diffuses with rate
$t^{(p+1)/d}$ and, up to a constant $A = -i \int_Rxf(x)dx$, it has $G'(x)$ as profile function.

Motivated by equation (\ref{aprox-formal}), we define the Renormalization Group operator as an operator acting in the space of initial conditions as follows
\begin{equation}
\label{eq:RG-oper}
\left(R_Lf\right)(x) =
L^{2(p+1)/d}
u\left(L^{(p+1)/d} x, L\right),
\end{equation}
where $L>1$ and $u$ is the solution to (\ref{eq:nao-lin}). Accordingly to the above definition, nonlinearities of the form $u^ju_x$,  $j\geq 1$, are irrelevant in the RG sense. Notice that, in \cite{bib:souz-brag-more-art,bib:souz-brag-more-marginal}, the function $u$ is rescaled by a factor of $L^{(p+1)/d}$ in the definition of the RG operator. The modification introduced in (\ref{eq:RG-oper}) is crucial because some terms in the nonlinearity (\ref{def:irre-pert}), which according to the old definition were classified as marginal, are now irrelevant, implying that the asymptotics of solutions to (\ref{eq:nao-lin}) is, basically, the one given by (\ref{aprox-formal}) and that is what we will prove in this paper. Also, as it will be shown in Lemma \ref{lem:pon:fix}
the profile function $G'(x)$ in (\ref{aprox-formal}) is a RG fixed point.

In order to state our results, we now introduce the space for the initial data. Since we are considering $f$ with zero mass, it is
necessary to incorporate the second derivative term in the definition below. Therefore, given $q > 1$, we consider
	\begin{equation}
		\label{def:esp-bq}
		\B_q=\left\{f\in L^1(\R): \widehat{f}(\omega) \in C^{2}(\R) \ \mbox{and} \ \|f\|_q<+ \infty \right\},
	\end{equation}
where $\|f\| = \sup_{\omega \in \R}(1+|\omega|^q)\left(|\widehat{f}(\omega)|+|\widehat{f}'(\omega)|+|\widehat{f}''(\omega)|\right)$.

\begin{theorem}
\label{teo:com-ass-full-nonl}
	Let $u(x,t)$ be the solution to (\ref{eq:nao-lin}) with $\lambda \in [-1,1]$, where $G(x,t)$ satisfies
the conditions ${\bf (G)}$ with $d>1$ and $q>3/2$, $f\in \B_q$ satisfies the zero mass hypothesis and $F$ is given by (\ref{def:irre-pert}). There exists positive constants $A$ and $\bar{\epsilon}$ such that, if $\|f\|<\bar{\epsilon}$, then
\begin{equation}
\label{cotadisc}
\lim_{t\rightarrow\infty}\|t^{2(p+1)/d}u(t^{(p+1)/d}.,t)-AG_p\|=0,
\end{equation}
where $A=A(d,f_0,F,p,q)$ and
\begin{equation}
		\label{def:G_p}
		G_p = G'\left(\cdot, \frac{1}{p+1}\right).
	\end{equation}
\end{theorem}

{\bf Remarks:} 1) Notice that property $(G_1)$ of $\textbf{(G)}$ is valid for $q>1$ and similarly, the definition of the $\B_q$
spaces also holds for $q>1$. The linear case will be treated under these conditions, see Theorem \ref{Teo:pri}. However, the restriction $q>3/2$ in the above theorem is necessary to ensure that the terms of the nonlinearity belong to $L^2$. 2) The above prefactor $A$, which we will show to be the limit of a sequence of prefactors $A_n$, can be found explicitly.

\section{\large{The Linear Case}}

In order to be able to fully discuss equation (\ref{eq:nao-lin}), we must present some properties of the kernel $G$ and consider first the employment of the RG
method to $u(x, t)$ given by the linear component of (\ref{eq:nao-lin}), that is, $u$ satisfies (\ref{eq:nao-lin}) with $\lambda=0$,
for $t>1$, $x\in\R$.
Throughout this paper, we consider $G$ satisfying $(G_3)$, $(G_2)$ for $d>1$ and $(G_1)$ for $q>3/2$ and the initial data $f$ in the space $\B_q$ for this particular $q$. Our goal in this section is to prove the following:
\begin{theorem}
\label{Teo:pri} Let $G$ be a kernel satisfying conditions {\bf (G)}, $f\in \B_{q}$ satisfying the zero mass condition, $A=-i\tra{f}'(0)$ and
$u$ solution to (\ref{eq:nao-lin}) with $\lambda=0$. Then,
\begin{equation}
\label{lim:pri}
\lim_{t\rightarrow \infty}\|t^\frac{2(p+1)}{d}u(t^\frac{p+1}{d}.,t)-AG_p\|=0.
\end{equation}
\end{theorem}

To prove Theorem \ref{Teo:pri} we will establish some
properties of the kernel $G$ which follow from conditions {{\bf (G)}}.
We denote $\widehat G'(\omega, t)\equiv \partial\widehat G/\partial \omega(\omega,t)$.
\begin{lema}
\label{Lem:Lim:G}
If $G(x,t):\R\times(0,\infty)\to \R$ satisfies property $(G_1)$, then, for $j=0,1,2$, $\widehat G^{(j)}(\omega,1)\in L^\infty(\R)$ and
$\sup_{\omega\in\R}(1+|\omega|^q)|\omega|^2|\tra G^{(j)}(\omega, 1)|<\infty.$
\end{lema}

{\bf{Proof: }}
It follows from property ($G_1$) and from Fourier Transform results.
\eop

We then denote throughout the paper:
\begin{equation}
	\label{cons:sup-G-j}
	K_j \equiv \sup_{\omega\in\R}|\tra G^{(j)}(\omega,1)| {\mbox{ and }} C_j\equiv\sup_w(1+|w|^q)|w|^2|\hat G^{(j)}(w,1)|
	\ \mbox{for} \ j = 0, 1, 2.
\end{equation}

It follows from Lemma \ref{Lem:Lim:G} and from property $(G_2)$ of $\textbf{(G)}$, that $\tra G(\omega,t)$ is well defined for $t>0$ and
\begin{equation}
	\label{tra G-j}
	\tra G^{(j)}(\omega,t)=t^{j/d}
	\partial_\omega^j\widehat{G}(t^{1/d}\omega,1),
	~~~~\mbox{ for } \ j = 0, 1, 2, \, \, \forall t>0, \  \omega\in \R.
\end{equation}
Also, condition ($G_3$) of $\textbf{(G)}$ in the Fourier space is
\begin{equation}
	\label{cond:1.3:tra G}
	\tra G(\omega,t)=\tra G(\omega,t-s)\tra G(\omega, s) ~~ t>s>0 \mbox{ and } \omega\in\R.
\end{equation}
These results together lead to the following lemmas:
\begin{lema}
\label{G: decres}
Suppose $G(x,t):\R\times(0,\infty)\to \R$ satisfies properties $(G_1)$, $(G_2)$ and $(G_3)$.
Given $t_1, t_2 \in (0,\infty)$, with $t_1<t_2$, then, for all $\omega\in\R$,
$$
|\widehat G(\omega,t_2)|\leq K_0|\widehat G(\omega,t_1)|,
$$
$$
|\widehat G'(\omega,t_2)|\leq K_1(t_2-t_1)^\frac{1}{d}|\widehat G(\omega,t_1)|+K_0|\widehat G'(\omega,t_1)|
$$
and
$$
|\tra G''(\omega,t_2)|\leq K_2(t_2-t_1)^\frac{2}{d}|\tra G(\omega,t_1)|+
		2K_1(t_2-t_1)^\frac{1}{d}|\tra G'(\omega,t_1)|+K_0|\tra G''(\omega,t_1)|
$$
with $K_j$, for $j = 0, 1, 2$, is given by (\ref{cons:sup-G-j}).
\end{lema}

\begin{lema}
\label{lem:tra G(0,t)}
If $G(x,t):\R\times(0,\infty)\to \R$ satisfies conditions $(G_1)$, $(G_2)$ and $(G_3)$, then, for all $t>0$,
$\int_{\R} G(x,t)dx=1.$
\end{lema}

\subsection{The RG Operator for the Integral Linear Equation}
\label{sec:op-rg-lin}

In Braga et al. \cite{bib:souz-brag-more-art}, we have established the RG method to integral equations. Unlike the case where we look for
the long-time behavior of solutions to PDEs, for integral equations, the definition of the RG operator is slightly more refined. In fact, for equation equation (\ref{eq:nao-lin}), with $\lambda =0$, not even the kernel
is scale invariant.
In order to define our RG transformation, let us consider $u$ given by equation (\ref{eq:nao-lin}), with $\lambda =0$, and define, for $t\in(1,L]$, $f_0\equiv f$,
$u_0(x,t)=\int G(x-y,s(t)) f_0(y)dy$ and, for $n=1, 2, \cdots$,
$$
u_{n}(x,t) \equiv \int{G(x-y,s(t)-s(L^n))u_{n-1}(y, L^n)dy}, \,\,\, t\in(L^n,L^{n+1}],
$$
and, for  $n = 0, 1, \cdots$,
$$
	u_{f_n}(x,t)\equiv\int{G\left(x-y,s_n(t)\right) f_n(y)}dy, \,\,\, t\in(1,L],
$$
where
\begin{equation}
\label{def:s_n(t)}
s_n(t)=\frac{s(L^nt)-s(L^n)}{L^{n(p+1)}}
\end{equation}
and
\begin{equation}
	\label{def:rg:lin}
	f_{n+1}\equiv R^0_{L,n}f_n \equiv L^{2(p+1)/d}u_{f_n}(L^{(p+1)/d}\cdot,L).
\end{equation}
With the above definitions, it is not hard to see that the RG operator satisfies the semigroup property, that is:
\begin{equation}
	\label{eq:prop-semi-grup}
R^0_{L^n,0}f_0=
(R^0_{L,n-1}\circ...\circ R^0_{L,1}\circ R^0_{L,0})f_0
\end{equation}
and so, the limit in (\ref{lim:pri}) will be obtained by studying the
dynamics of operators $R^0_{L,n}f_n$, $n=0, 1, 2, \cdots$ in the initial data space $\B_q$.

It follows from definition (\ref{def:s(t):int}) that we can write the function $s(t)$ as
\begin{equation}
\label{def:r(t)}
s(t) =  \frac{t^{p+1}-1}{p+1}+r(t),
\end{equation}
where $r(t)=o(t^{p+1})$. Therefore, there is $L_0>1$ such that, for $L>L_0$,
$|r(L)|/L^{p+1}<1/[4(p+1)]$. Furthermore, from (\ref{def:s_n(t)}), we can write
$$
	s_n(t)\equiv\frac{t^{p+1}-1}{p+1}+r_n(t), \,\, {\mbox{ where }} \,\,\ r_n(t)=\frac{r(L^nt)-r(L^n)}{L^{n(p+1)}}.
$$
Then, for $L>L_0>1$, $|r_n(L)|<1/[2(p+1)]$.
Defining
\begin{equation}
	\label{def:L1}
	L_1\equiv \max \{L_0,3^{1/(p+1)}\}
\end{equation}
we get, for $L>L_1$,
	\begin{equation}
		\label{cot:s_n(L)}
		\frac{1}{6(p+1)}<\frac{s_n(L)}{L^{p+1}}<\frac{3}{2(p+1)},
	\end{equation}
for all $n\geq 0$. Furthermore, from the properties of $G$ and the Fourier transform, 
$R^0_{L,n}g \in \B_q $, for all $g \in \B_q$ and $n=0, 1, \cdots$. In the next lemma, we will show that $G_p(x)$, that is, the derivative with respect to the variable $x$ of the function $G(x,t)$ at $t=1/(p+1)$ is also in $\B_{q}$.

\begin{lema}
	\label{lem:cot:G_p}
There is a constant $C_{d,p,q}$ such that $\|G_p\|\leq C_{d,p,q}.$
\end{lema}
{\bf{Proof: }}
It follows from (\ref{def:G_p}) and (\ref{tra G-j}), with $t = \frac{1}{p+1}$ and $j \in \{0,1, 2\}$, that
$$
\|G_p\|\leq \sup_{\omega \in \R}(1+|\omega|^{q})(2+|w|)
\left[
\sum_{j=0}^{2}
\left(\frac{1}{p+1}\right)^{j/d}\Bigg|
\widehat{G_p}^{(j)}\left(\left(\frac{1}{p+1}\right)^{\frac{1}{d}}\omega,1\right)\Bigg|
\right].
$$
Using Lemma \ref{Lem:Lim:G} and taking $k=[1/(p+1)]^{\frac{1}{d}}\omega$ and the supreme in $k$, we get the result, with
$C_{d,p,q}=2(p+1)^{\frac{{(q+1)}}{d}}\sup_{k \in \R}(1+|k|^{q})(1+|k|)\sum_{j=0}^{2}|\tra G^{(j)}(k,1)|$.
\eop

From now on, $L_1$ will always denotes the constant given by (\ref{def:L1}). In the next lemma we prove that $R^0_{L^n}G_p \in \B_q$:

\begin{lema}
	\label{lem:G_p}
	There is a constant $\tilde{K}$ such that $\|R^0_{L^n}G_p\|\leq \tilde{K}$ for all $L>L_1$.
\end{lema}
\textbf{Proof: } It follows from (\ref{def:G_p}), (\ref{cond:1.3:tra G}) and the properties of the Fourier transform that
\begin{equation}
	\label{eq:RLfp}
	\F[R_{L^n}^0G_p](\omega)=i\omega\tra G\left(\frac{\omega}{L^{n(p+1)/d}}, \frac{1}{p+1}+s_0(L^n)\right).
\end{equation}
Taking the derivatives of the above equation and using (\ref{tra G-j}) for $j=0, 1 \mbox{ and } 2$ and
(\ref{cot:s_n(L)}), it follows from Lemma \ref{G: decres} that
\begin{eqnarray*}
	\|R_{L^n}^0G_p\| &\leq &
\sup_{\omega \in \R}(1+|\omega|^{{q}})(|\omega|+1)
\left[2K_0+2K_1\left(\frac{7}{3(p+1)}\right)^{\frac{1}{d}}+K_2\left(\frac{7}{3(p+1)}\right)^{\frac{2}{d}}\right]\nonumber \\
	& & \Bigg[\Bigg|\tra G\left(\omega,\frac{1}{6(p+1)}\right)\Bigg|+
	\Bigg|\tra G'\left(\omega,\frac{1}{6(p+1)}\right)\Bigg|+\Bigg|\tra G''\left(\omega,\frac{1}{6(p+1)}\right)\Bigg|\Bigg].
\end{eqnarray*}


Applying the change of variable $k=[6(p+1)]^{-1/d}\omega$, we get, as an upper bound for $\|R_{L^n}^0G_p\|$:
\begin{eqnarray*}
	\sup_{k\in \R}& & [6(p+1)]^{\frac{{q+1}}{d}}(1+|k|^{q})(1+|k|)
\left[2K_0+2K_1\left(\frac{7}{3(p+1)}\right)^{\frac{1}{d}}+K_2\left(\frac{7}{3(p+1)}\right)^{\frac{2}{d}}\right]\nonumber \\
	& & \left\{|\tra G\left(k,1\right)|+|\tra G'\left(k,1\right)|+|\tra G''\left(k,1\right)|\right\}.
\end{eqnarray*}
From Lemma \ref{Lem:Lim:G}, the right hand side of the above inequality is finite, which finishes the proof.
\eop

In the next lemma we show that $G_p$ is an asymptotic fixed point for the RG operator $R^0_{L^n}G_p$, that is, $R^0_{L^n}G_p\to G_p$ when $n\rightarrow \infty$.

\begin{lema}
	\label{lem:pon:fix}
	There are positive constants $M$ and $N$, depending on $p, q, d$, such that
	\begin{equation}
		\label{equ:pon:fix}
		\|R_{L^n}^0G_p-G_p\|\leq M\left|\frac{r(L^n)}{L^{n(p+1)}}\right|^{\frac{1}{d}}
		+N\left|\frac{r(L^n)}{L^{n(p+1)}}\right|^{\frac{2}{d}}.
	\end{equation}
\end{lema}
\textbf{Proof: } It follows from (\ref{eq:RLfp}), (\ref{cond:1.3:tra G}) and (\ref{tra G-j}) with $j = 0$, that
$$
	\F[R_{L^n}^0G_p-G_p](\omega)=
	i\omega\tra G\left(\omega,\frac{1}{p+1}\right)\left[\tra G\left(\left(\frac{r(L^n)}{L^{n(p+1)}}\right)^{\frac{1}{d}}\omega,1\right)-1\right].
$$
From Lemmas \ref{Lem:Lim:G} and \ref{lem:tra G(0,t)} and from the Mean Value Theorem, we conclude that
$$
	\left|\tra G\left(\omega\left(\frac{r(L^n)}{L^{n(p+1)}}\right)^{\frac{1}{d}},1\right)-1\right|\leq K_1\left|\frac{r(L^n)}{L^{n(p+1)}}\right|^{\frac{1}{d}}|\omega|.
$$
Using the above results, we bound $|\F[R_{L^n}^0G_p-G_p]^{(j)}(\omega)|$, for $j=0,1,2$, respectively by
$$
	K_1 \omega^2\left|\frac{r(L^n)}{L^{n(p+1)}}\right|^{\frac{1}{d}}\left|\tra G\left(\omega,\frac{1}{p+1}\right)\right|,
$$
$$
\left|\frac{r(L^n)}{L^{n(p+1)}}\right|^{\frac{1}{d}}
\left[
2K_1\left|\omega \right|
\left|\tra G\left(\omega,\frac{1}{p+1}\right)\right|
+K_1\omega^2
\left|\tra G'\left(\omega,\frac{1}{p+1}\right)\right| \right]
$$
and
$$
K_1 \left|\frac{r(L^n)}{L^{n(p+1)}}\right|^{\frac{1}{d}}\left[
4|\omega| \left|\tra G' \left(\omega,\frac{1}{p+1}\right)\right| +
\omega^2\left|\tra G'' \left(\omega,\frac{1}{p+1}\right)\right| +
2\left|\tra G \left(\omega,\frac{1}{p+1}\right)\right|
\right]
$$
$$
+
K_2|\omega|\left|\frac{r(L^n)}{L^{n(p+1)}}\right|^{\frac{2}{d}}
\left|\tra G \left(\omega,\frac{1}{p+1}\right)\right|.
$$
Finally, the result follows from the change of variable $y=1/(p+1)^\frac{1}{d}\omega$ and from the definitions:
$M=K_1(M_1+M_2+M_3)$, where $M_1=(p+1)^{\frac{q+2}{d}}\sup_{y \in \R}(1+|y|^q)
y^2\left(|\tra G(y,1)|+|\tra G'(y,1)|+|\tra G''(y,1)|\right)$,
$M_2=2(p+1)^{\frac{q+1}{d}}\sup_{y \in \R}(1+|y|^q)|y|\left(|\tra G(y,1)|+2|\tra G'(y,1)|\right)$,
$M_3=2(p+1)^{\frac{q}{d}}\sup_{y \in \R}(1+|y|^q)|\tra G(y,1)|$ and
$N = K_2(p+1)^{\frac{q+1}{d}}\sup_{y \in \R}(1+|y|^q)|y||\tra G(y,1)|.$
\eop
%
%

The next lemma states that, for $L$ sufficiently large, $R^0_{L,n}$ is a contraction when acting on the space of functions
$g\in \B_q$ with zero mass and null first momentum ($\tra  g (0) = \tra g' (0) =0$).

\begin{lema}(Contraction Lemma)
	\label{lemadacontracao} Given $q > 1$, if $g\in \B_q$ is such that $\tra g(0)= \tra g'(0)=0$, then, there is a constant $C=C(d,p,q)>0$ such that
	\begin{equation}
		\label{lema:contr}
		\|R^0_{L,n}g\|\leq \frac{C}{L^{(p+1)/d}}\|g\|,~~\forall~L>L_1.
	\end{equation}
\end{lema}

The proof of the above theorem follows basically from the Fundamental Theorem of
Calculus, the definition of the $\B_q$ space and the properties of the Fourier Transform.

The method in order to obtain Theorem \ref{Teo:pri} consists in decomposing the initial data $f$ into two components, the first a multiple of $R^0_{L,n}G_p$. We will use the next lemma to conclude that, as the second component of this decomposition has zero mass and null first momentum,
its norm goes to zero when $n\to\infty$.
It will follow from this fact and from Lemma \ref{lem:pon:fix} that the function $u(x,t)$,
defined by (\ref{eq:nao-lin}), with $\lambda = 0$, and properly rescaled,
will behave, when $t \to \infty$, as a multiple of $G_p$.

\begin{lema}
	\label{cot:g_n}
	Let $f\in \B_q$, $f_0\equiv f$, $A\equiv -i\tra f_0'(0)$ and $f_n=R^0_{L,n-1}f_{n-1}$,
	$n=1, 2, \cdots$. Given $L>L_1$, there are functions $g_n\in \B_q$, $n=0,1,2,...$, such that $\tra g_n(0)=\tra g_n'(0)=0$, $\forall \,\, n,$
	and
	\begin{equation}
		\label{def:g_n}
		f_0=AG_p+g_0,~~~~~~~~~~f_n=AR^0_{L^n}G_p+ g_n.
	\end{equation}
Furthermore,
	\begin{equation}
		\label{nor:g_n}
		\|g_n\|\leq \left(\frac{C}{L^{(p+1)/d}}\right)^n\|g_0\|,\,\,\, n=0, 1, 2, \cdots
	\end{equation}
	where $C$ is the constant from Lemma \ref{lemadacontracao}.
\end{lema}

The proof of Lemma \ref{cot:g_n} follows directly from induction on $n$ and we are finally ready to prove Theorem \ref{Teo:pri} and then analyze the nonlinear case.

\textbf{Proof of Theorem \ref{Teo:pri}: }
Let $C$ be the constant from Lemma \ref{lemadacontracao} and define
	$L_2\equiv \max\{L_1,C^{d/(p+1)}\}$.
If $L>L_2$, it follows from (\ref{equ:pon:fix}), (\ref{def:g_n}) and (\ref{nor:g_n}) that
$$
	\|L^{2n(p+1)/d}u(L^{n(p+1)/d}\cdot ,L^n)-AG_p\|\leq \left(\frac{C}{L^{(p+1)/d}}\right)^n\|g_0\|+M|A|\Big|\frac{r(L^n)}{L^{n(p+1)}}\Big|^\frac{1}{d}.
$$
Given $\delta \in (0,1)$, we take $L_3>L_2$ such that $L_3^{\delta(p+1)/d}>C$. Then, for $L>L_3$,
$(CL^{-(p+1)/d})^n\leq L^{-n(p+1)(1-\delta)/d}$ and, for $t=L^n$ with $L>L_3$,
\begin{equation}
	\label{nor:final2}
	\|	t^{2(p+1)/d}u(t^{(p+1)/d}.,t)-AG_p\|\leq\frac{\|g_0\|}{t^{(p+1)(1-\delta)/d}}+M|A|\left|\frac{r(t)}{t^{p+1}}\right|^{1/d}.
\end{equation}
To extend this result, we just replace $L$ with $\tau^{\frac{1}{n}}L$ in the estimates, where $\tau \in [1,L]$ and $L > L_3$. Since the constants are independent from $L$, the above inequality holds for $t =\tau L^n$.
\eop
\section{\large{The Nonlinear Case}}

From now on, we will consider the nonlinear equation (\ref{eq:nao-lin}), with $F$ given by (\ref{def:irre-pert}). In Section \ref{subsec1}, we will prove that the integral equation (\ref{eq:nao-lin}) has a unique local solution and use this, in Section \ref{lem:renom}, to obtain our main result, which is Theorem \ref{teo:com-ass-full-nonl}.

\subsection{Local existence and uniqueness of the solution}
\label{subsec1}

Our goal in this section is to prove that there is a unique local solution to the nonlinear integral equation (\ref{eq:nao-lin}). In order to do that,
let $u_f$ be the solution to the linear integral equation (\ref{eq:nao-lin}), with $\lambda = 0$, and, given $L,q>1$, define
\begin{eqnarray}
\label{def:BL}
B^{(L)}=\{u:\mathbb{R} \longrightarrow \mathbb{R} \times[1,L] ; \ u(\cdot, t) \in \B_q, \ \forall t \in [1,L] \}
\end{eqnarray}
equipped with the norm $\|u\|_L=\sup_{t\in [1,L]}\|u(\cdot, t)\|$,
and
\begin{equation}
\label{def:Bf}
B_{f}\equiv\{u\in B^{(L)}:\|u-u_{f}\|_L\leq \|f\|\}.
\end{equation}
Furthermore, consider the operator
	$T(u)\equiv u_{f}+N(u)$,
 where
\begin{equation}
	\label{def:N(u)}
	N(u)(x,t)=\lambda\int_1^{t}\int {G(x-y,s(t)-s(\tau))F(u,u_x)(y,\tau)dy d\tau}.
\end{equation}
We will prove the following:
\begin{theorem}
\label{teo:exis-uni-local}
Given $d>1$, $q>3/2$, $L>L_1$ and $\lambda \in [-1,1]$, there exists $\epsilon>0$ such that, if $\|f\|<\epsilon$, then the integral equation (\ref{eq:nao-lin}) has a unique solution in $B_f$.
\end{theorem}

{\bf{Proof: }} Notice that $N(u)(x,t)=\lambda \sum_{j\geq \alpha+1}H_{j}(u)(x,t)$, where
$$
H_{j}(u)(x,t) = c_{j}\int_0^{t-1}\int {G(x-y,\phi (\tau))\left(u^{j}\right)_x(y,t-\tau)dy d\tau},
$$
$\phi (\tau) = s(t)-s(t-\tau)$ and $c_{j}=a_{j-1}/j$, with $a_j$ being the coefficient in (\ref{def:irre-pert}).
First we point out that, from Lemma \ref{Lem:Lim:G} and (\ref{tra G-j}), if $|w|>1$, then, $|\phi(\tau)|^{1/d}|w||\hat G^{(j)}\left(\phi(\tau)|^{1/d}w,1\right)|\leq C_j$, where $C_j$ is given by (\ref{cons:sup-G-j}). Furthermore, since $\phi(\tau)\sim t^p\tau$, we get that $\int_0^{t-1} [\phi(\tau)]^{-1/d}d\tau \equiv C_{t,d}<\infty$, if $d>1$. Also, $|u(x,t)|\leq C_q$ and $\|u\|_L \leq \bar{C}_L\|f\|$, for all $u \in B_f$, where $C_q=(2\pi)^{-1}\int[1+|w|^q]^{-1}dw$ and $\bar{C}_L= 1+K_0+2K_1\left(s(L)\right)^{\frac{1}{d}}+K_2\left(s(L)\right)^{\frac{2}{d}}$.
Let $C=\left(2^{q+1}+3\right)\int_{\R}[1+|x|^q]^{-1}dx$ and
$S_1(z)= \sum_{j\geq \alpha +1}\left(\frac{C}{2\pi}\right)^{j-1}|c_{j}|z^{j-2}$. By defining $\rho = \min \left\{r C_q^{-1}, 2\pi rC^{-1} \right\}$, we guarantee that the series $S_1(z)$ is convergent and that the operators $T$ and $N$ are well defined. Using (\ref{cons:sup-G-j}), (\ref{tra G-j}) and the definition of $\B_q$ and $B^{(L)}$ spaces, it follows that
$\|N(u)\|_L \leq B_{L,d}\bar{C}^2_LS_1(\rho)\|f\|^2$, with
$$
B_{L,d}\equiv \left[9K_0+3C_1+[s(L)]^{1/d}(7K_1+C_2)+[s(L)]^{2/d}K_2\right](L-1)+3C_0 C_{L,d}.
$$
Taking $\epsilon_1 \equiv \min \left\{[B_{L,d}\bar{C}^2_LS_1(\rho)]^{-1}, \rho\bar{C}_L^{-1}\right\}$ and $\|f\|<\epsilon_1$, allows us to ensure that $T$ maps $B_f$ into itself. In a similar way, we can determine the condition for $T$ to be a contraction.  Indeed, proceeding as before, we obtain, as an upper bound for $\|N(u)-N(v)\|_L$,
$$
B_{L,d}\left(\|u\|_L^{j -1}+ \|u\|_L^{j -2}\|v\|_L + \cdots +\|v\|_L^{j -1} \right)\|u-v\|_L\dsum_{j\geq \alpha+1}|c_j|\left(\frac{C}{2\pi}\right)^{j-1}.
$$
Since $j\geq \alpha+1\geq 2$, we can bound one term in each summand within the parentheses above by $\bar{C}_L\|f\|$, and thus we obtain
$\|N(u)-N(v)\|_L \leq B_{L,d}\bar{C_L}S_2(\rho)\|f\| \|u-v\|_L$, where
$S_2(z)=\sum_{j\geq\alpha+1}|c_j|\left(\frac{C}{2\pi}\right)^{j-1}jz^{j-2}$. Accordingly, the theorem is established, provided
$\epsilon \equiv \min\{\epsilon_1,[2B_{L,d}\bar{C_L}S_2(\rho)]^{-1}\}$.
\eop

\subsection{Renormalization}
\label{lem:renom}

In order to obtain the asymptotic behavior of solutions using the renormalization group approach, one must analyze the existence and stability of fixed points of an appropriate RG operator. Once such operator has been found for a particular problem, the method is iterative and the application
of the RG transformation progressively evolves the solution in time and at the same time renormalizes
the terms of the equation.

For each defined RG operator, we classify the nonlinearities of the problem according to their importance in the asymptotic behavior of the solution. This classification in the case of initial value problems is quite straightforward, with just a substitution of the renormalized solution to the equation. However, in the case of the integral equation, it is necessary to explore the properties of the kernel of the equation to obtain the classification, as we will show below.
Let us consider equation (\ref{eq:nao-lin}) with $F(u,u_x)$ given by (\ref{def:irre-pert}) and, given $L>1$, define
\begin{equation}
\label{def:un:cas:irr}
u_n(x,t)\equiv L^{2n(p+1)/d}u(L^{n(p+1)/d}x,L^nt), ~~ t\in [1, L], ~~ n=0, 1, 2, \cdots.
\end{equation}
We will prove that $u_n(x,t)$ satisfies the renormalized equation
$$
u_{n}(x,t)= \int G(x-y,s_n(t))f_n(y)dy +
$$
\begin{equation}
\label{equ:ren:nao:lin:fin}
\lambda_n\int_1^t\int G(x-y,s_n(t)-s_n(\tau))F_{L,n}(u_{n},\partial_x u_n)(y,\tau)dyd\tau,
\end{equation}
where $\lambda_n$ is the {\em generalized coupling constant}
\begin{equation}
\label{def:lambdan}
\lambda_n=L^{-\frac{n}{d}[(2\alpha+3)(p+1)-2(p+1)-d]}\lambda,
\end{equation}
\begin{equation}
\label{def:Fln}
F_{L,n}(u_n,\partial_x u_n)=\dsum_{j\geq \alpha} a_j L^{\frac{2n}{d}(p+1)(\alpha-j)}u_n^j \partial_x u_n
\end{equation}
and $f_n$ is the {\em renormalized initial data}
\begin{equation}
\label{def:fn}
f_n(x)\equiv L^{2n(p+1)/d}u(L^{n(p+1)/d}x,L^n).
\end{equation}
In fact, if we write $u_n(x,t)=a(x,t)+b(x,t)$ where
$$
a(x,t)\equiv L^{2n(p+1)/d}\int {G(L^{n(p+1)/d}x-y, s(L^nt))f(y)dy}\,\,\, +
$$
$$\lambda L^{2n(p+1)/d}\int_1^{L^n}\int {G(L^{n(p+1)/d}x-y,s(L^nt)-s(\tau))F(u,u_x)(y,\tau)dy d\tau}
$$
and
$$b(x,t) \equiv \lambda L^{2n(p+1)/d}\int_{L^n}^{L^nt}\int {G(L^{n(p+1)/d}x-y,s(L^nt)-s(\tau))F(u,u_x)(y,\tau)dy d\tau},
$$
it follows from conditions $(G_2)$ and $(G_3)$ that $a(x,t)=\int{G(x-\omega,s_n(t))f_n(\omega)d\omega}$, where $s_n$ and $f_n$ are given, respectively by (\ref{def:s_n(t)}) and (\ref{def:fn}).
Now if we consider the change of variables $y=L^{n(p+1)/d}\omega$, $\tau=L^nq$ and we divide and multiply the definition of $b(x,t)$ by $L^{n(2\alpha+3)(p+1)/d}$, then, it follows from condition $(G_2)$ that $$L^{n(p+1)/d}G(L^{n(p+1)/d}(x-\omega),s(L^nt)-s(L^nq))=G(x-\omega,s_n(t)-s_n(q)).$$
Therefore, $b(x,t)$ can be written as the second term of the equation (\ref{equ:ren:nao:lin:fin}), where $\lambda_n$ and $F_{L,n}$ are given, respectively by (\ref{def:lambdan}) and (\ref{def:Fln}), which concludes the proof.

We are then able to classify the nonlinearity $F(u, u_x)$  into its universality class, according to its asymptotic behavior. Defining $d_F\equiv (2\alpha+3)(p+1)-2(p+1)-d$, relevant (or subcritical) and marginal (or critical) nonlinearities are obtained when $d_F<0$ and $d_F=0$, respectively. In this context, we are interested in considering irrelevant (or supercritical) perturbations, namely those for which $d_F>0$, or, more precisely, for which $2\alpha +3>2+d/(p+1)$.

From now on we assume $d_F>0$, that is, we consider $\alpha>\alpha_c$, where $\alpha_c$ is defined in~(\ref{def:alpha-c}). Following the classification of perturbations, we obtain the renormalization step that ensures the iterative applicability of the procedure. Given $L>1$, consider $B^{(L)}$ and $f_n$ given, respectively, by (\ref{def:BL}) and
(\ref{def:fn}), and define
$B_{f_n}\equiv\{u_n\in B^{(L)}:\|u_n-u_{f_n}\|\leq \|f_n\|\}$ and 
$T_n(u_n)\equiv u_{f_n}+N_n(u_n)$,
where $u_{f_n}$ is the solution to the linear integral equation, that is, (\ref{equ:ren:nao:lin:fin}) with $\lambda_n=0$, and
\begin{equation}
\label{def:Nn(u)}
N_n(u_n)(x,t)=\lambda_n\int_1^{t}\int {G(x-y,s_n(t)-s_n(\tau))F_{L,n}(u_n,\partial_x u_n)(y,\tau)}dy d\tau
\end{equation}
where $s_n(t)$, $\lambda_n$ and $F_{L,n}(u_n)$ are respectively given by (\ref{def:s_n(t)}),
(\ref{def:lambdan}) and (\ref{def:Fln}). To simplify the notation, we define $\nu_{n}(x)\equiv N_n(u_n)(x,L)$. The next lemma is necessary to ensure that the operators above are well-defined.
\begin{lema}
\label{lema:existn}
Given $n\in \N$ and $L > L_1$, there exists $\epsilon_n > 0$ such that, if $\|f_n\| < \epsilon_n$,
then the integral equation (\ref{equ:ren:nao:lin:fin}) has a unique solution in $B_{f_n}$.
\end{lema}
{\bf{Proof: }} First we notice that, defining $\phi_n(\tau)=s_n(t)-s_n(t-\tau)$ we get, for $L>L_1$,
$[\phi_n(\tau)]^{1/d}\leq [s_n(t)]^{1/d}\leq \left(\frac{3L^{p+1}}{2(p+1)}\right)^{1/d}$ and also, since $\phi_n(\tau)\sim \frac{t^{p+1}-(t-\tau)^{p+1}}{p+1}+o(1)$, there exists $\xi \in (t-\tau,t)$ such that $\phi_n(\tau)\sim \xi^p\tau+o(1)$ and, therefore,
$\int_0^{t-1}[\phi_n(\tau)]^{-1/d}d\tau \leq \bar{C}_{t,d}$, where $\bar{C}_{t,d}$ is a constant depending only on $t$ and $d$. We also notice that $\|u_n\|_L\leq \bar{C}_{L,n}\|f_n\|$, where $\bar{C}_{L,n}= 1+K_0+2K_1\left(s_n(L)\right)^{\frac{1}{d}}+K_2\left(s_n(L)\right)^{\frac{2}{d}}$. Then, proceeding as in the proof of Theorem \ref{teo:exis-uni-local} and recalling that $d_F\equiv (2\alpha+3)(p+1)-2(p+1)-d$, we get:
$$
\|N_n(u_n)\|_L \leq L^{-\frac{nd_F}{d}}Q_n \|f_n\|^2 {\mbox{ and }} \|N_n(u_n)-N_n(v_n)\|_L \leq L^{-\frac{n d_F}{d}}Q_n \|f_n\|\|u_n-v_n\|_L,
$$
where $Q_n$ is a constant depending on $L$, $n$, $d$ and $p$, given by
$\left[\left(9K_0+3C_1+[s_n(L)]^{\frac{1}{d}}(7K_1+C_2)+[s_n(L)]^{\frac{2}{d}}K_2\right)(L-1)+3C_0 \bar{C}_{L,d}\right]\bar{C}^2_{L,n}S_2(\rho)$ and $S_2(z)$ is defined as in the proof of Theorem \ref{teo:exis-uni-local}.
We get the result by taking $\epsilon_n \equiv \min\{(2Q_n)^{-1},\rho (\bar{C}_{L,n})^{-1}\}$.
\eop

The above lemma guarantees that, if $\|f_n\|<\epsilon_n$, then, the
renormalized integral equation (\ref{equ:ren:nao:lin:fin}) has a unique solution that can be written in time $t=L$ as
$u_{n}(x,L)=u_{f_n}(x,L)+\nu_{n}(x)$.
It follows that, for every positive integer $n$, the renormalization group operator for the integral equation (\ref{equ:ren:nao:lin:fin}) is well defined:
\begin{eqnarray}
\label{def:rg:nao:lin}
(R_{L,n}f_{n})(x) \equiv L^{2(p+1)/d}u_n(L^{(p+1)/d}x,L),
\end{eqnarray}
which leads to the definition
\begin{eqnarray}
\label{def:f_n}
f_0=f \ \mbox{and} \ f_{n+1} = R_{L,n}f_{n}, \ \mbox{for} \ n =0, \ 1, \ 2, \ \cdots
\end{eqnarray}

Using definitions
(\ref{def:rg:nao:lin}) and (\ref{def:f_n}) we get that
\begin{equation}
\label{repre:nu}
f_{n+1}=R^0_{L,n}f_{n} + L^{2(p+1)/d}\nu_{n}(L^{(p+1)/d}\cdot),
\end{equation}
where $R^0_{L,n}$ is the linear operator RG given by (\ref{def:rg:lin}). Before proceeding to the renormalization, we first notice that, for $L>L_1$, from (\ref{cot:s_n(L)}), the constants $\bar{C}_{L,n}$ and $Q_n$ in the proof of Lemma \ref{lema:existn} can be uniformly bounded above by $\tilde{C}$ and $\tilde{Q}$, respectively. Therefore, defining
$
\sigma \equiv \min\{(2\tilde Q)^{-1},\rho/\tilde{C}\},
$
we get $\sigma <\epsilon_n$, for all $n$.

\begin{lema}(Renormalization Lemma)
	\label{lema:renorm}
Given $k\in \N$, $d>1$, $\alpha >\alpha_c$ and $L>L_1$, consider the integral renormalized equation (\ref{equ:ren:nao:lin:fin}), where $f_n$ is well defined, given by (\ref{def:f_n}), for $n=0,1,\cdots,k+1$ and $f_0$ has zero mass. Then, there are constants $A_n$ and functions $g_n \in \B_q$ such that, for $n=0,1,\cdots,k$, $\widehat{g_n}(0)=\widehat{g_n}'(0)=0$,
	\begin{equation}
	\label{eq:decomp}
	f_0=A_0G_p+g_0, \,\,\, 	f_{n+1}=A_{n+1}R^0_{L^{n+1}}G_p+g_{n+1},
	\end{equation}
\begin{equation}
\label{desigAn}
|A_{n+1}-A_n|\leq L^{-\frac{nd_F}{d}}\tilde{Q} \|f_n\|^2,
\end{equation}
\begin{equation}
\label{cota:gn}
\|g_{n+1}\|\leq \frac{C}{L^{(p+1)/d}}\|g_{n}\| +\tilde M L^{-\frac{nd_F}{d}}\|f_n\|^2,
\end{equation}
where $\tilde M  \equiv (L^{(q+1)(p+1)/d}+\tilde{K})\tilde{Q}$. In particular, $f_{n+1}$ has zero mass for $n=0,1,\cdots,k$.
\end{lema}
\textbf{Proof:} The argument follows from induction on $n$. Let $A_0= -i\tra f_0'(0)$ and define $g_0=f_0 -A_0 G_p$. Then, since $\hat G_p(0)=0$ and $\hat G'_p(0)=i$, we get that $\hat g_0(0)=\hat g_0'(0)=0$. By hypothesis, $f_1$ is well defined by $R_{L,0}f_0$ and, using (\ref{repre:nu}) and the decomposition for $f_0$,
$f_{1}=A_1R^0_{L}G_p+g_1$,
with $A_{1}\equiv A_{0}-i\widehat{\nu}'_{0}(0)$ and $g_1\equiv
R^0_Lg_0+L^{2(p+1)/d}\nu_{0}(L^{(p+1)/d}\cdot)+i\widehat{\nu}'_{0}(0)R^0_LG_p$. It follows from
the definition of $R_L^0$ and $\nu_0$ and, from (\ref{eq:RLfp}), that $\F(R^0_Lg_0)(0)=\F(R^0_{L}G_p)(0)=\hat\nu_0(0)=0$ and therefore, $\widehat g_1(0)=0$. Furthermore, since $[\F(R^0_{L}G_p)]'(0)=i$, we get also $\widehat{g_1}'(0)=0$, which proves (\ref{eq:decomp}) for $n=0$.
Now suppose $f_{n}$ is well defined for $n=0,1,...k$ and (\ref{eq:decomp}) holds with $\widehat{g_n}(0)=\widehat{g_n}'(0)=0$, for $n=0,1,\cdots,k-1$. Then, using (\ref{repre:nu}) and (\ref{eq:prop-semi-grup}), we get that $f_{k+1}=A_{k+1}R_{L^{k+1}}^0f_p^*+g_{k+1}$, with
$A_{k+1}=A_k-\widehat\nu_k'(0)$ and
\begin{equation}
\label{def:g_{j+1}}
g_{k+1}= R^0_{L,k}g_k+L^{2(p+1)/d}\nu_k(L^{(p+1)/d} \cdot)+i\widehat\nu_k'(0)R^0_{L^{k+1}}G_p,
\end{equation}
with $\widehat g_{k+1}(0)=\widehat g_{k+1}'(0)=0$. This also implies that $\hat f_{k+1}(0)=0$. Inequality (\ref{desigAn}) is straightforward from the fact that $L>L_1$ and from the bound for $\|N_n(u_n)\|_L$ in the proof of Lemma \ref{lema:existn}, since $ |A_{n+1}-A_n|=|\widehat
\nu_n'(0)|\leq \|N(u_n)\|_L$ and $Q_n \leq \tilde Q$, for all $n$. Inequality (\ref{cota:gn}) follows also from the bound for $\|N_n(u_n)\|_L$ and from lemmas \ref{lem:G_p} and \ref{lemadacontracao}.
\eop

Given $\alpha>\alpha_c$, consider $\delta \in (0,1)$ such that
\begin{equation}
\label{condicao-alpha}
(1-\delta)(p+1)<d_F
\end{equation}
and define
\begin{equation}
\label{def:L-delta}
L_{\delta}=\max\left\{L_1,[2C (1+C_{d,q,p})]^{\frac{d}{\delta (p+1)}}\right\}
\end{equation}
and
\begin{equation}
\label{def:D}
D\equiv 1 + \tilde{K}\sum_{j=0}^{\infty}\frac{1}{L^{j(p+1)(1-\delta)/d}}.
\end{equation}
where $C$, $C_{d,q,p}$ and $\tilde K$ are the constants given, respectively, in Lemmas \ref{lemadacontracao}, \ref{lem:cot:G_p} and \ref{lem:G_p}. Throughout the rest of this paper, $L_{\delta}$ will be referring to the constant above. The results of the next lemma ensure that all renormalized equations admit a unique solution, thereby guaranteeing that the algorithm's iterations are well defined.

\begin{lema}
\label{lem:con:g:f}
Consider $\delta \in (0,1)$ satisfying (\ref{condicao-alpha}) and let $L>L_{\delta}$. Then, there is $\bar\epsilon > 0$ such that, if $\|f_{0}\|<\bar\epsilon$, then
$f_{n}$ and $g_n$ given respectively by (\ref{def:fn}) and (\ref{eq:decomp}) are well defined for all $n\geq 1$ and satisfy
\begin{equation}
\label{con:fn}
\|f_{n}\|\leq D\|f_0\|
\end{equation}
and
\begin{equation}
\label{equ:gn}
\|g_{n}\|\leq\frac{\|f_0\|}{L^{n(p+1)(1-\delta)/d}}.
\end{equation}
\end{lema}

{\bf Proof:} Define
$\bar{\epsilon} \equiv
\min \left\{[2L^{(p+1)(1-\delta)/d}\tilde{M}D^2]^{-1},\sigma D^{-1}\right\}$ and assume $\|f_{0}\|<\bar\epsilon$. Inequalities (\ref{con:fn}) and (\ref{equ:gn}) follow inductively from Lemma \ref{lema:renorm} (see the proof of Lemma 33 in \cite{bib:souz-brag-more-art}).
\eop

{\bf Proof of Theorem \ref{teo:com-ass-full-nonl}:}
If $\|f_0\|<\bar\epsilon$, it follows from (\ref{desigAn}) that $|A_{n+1}-A_n|\leq L^{-[nd_F+(p+1)(1-\delta)]/d}\|f_0\|/2$ and, therefore, $A_n \to A$. From the Renormalization Lemma and from the semigroup property of the RG operator,
$$
\|L^{2n(p+1)/d}u(L^{n(p+1)/d}.,L^{n})-AG_p\|\leq |A|\|R^0_{L^n}G_p-G_p\|+\|g_n\|+|A_n-A|\|R^0_{L^n}G_p\|,
$$
which, from Lemmas \ref{lem:G_p}, \ref{lem:pon:fix} and from (\ref{equ:gn}) goes to zero as $n\to \infty$. In addition, we are able to provide an estimate for the rate of convergence, since the following upper bound can be established for $\|L^{2n(p+1)/d}u(L^{n(p+1)/d}.,L^{n})-AG_p\|$:
$$
|A| M\Big|\frac{r(L^n)}{L^{n(p+1)}}\Big|^{\frac{1}{d}}+|A| N\Big|\frac{r(L^n)}{L^{n(p+1)}}\Big|^{\frac{2}{d}}+\frac{\|f_0\|}{L^{n(p+1)(1-\delta)/d}}+
\frac{L^{-n[d_F+(p+1)(1-\delta)]/d}}{2(1-L^{-d_F/d})}\tilde K\|f_0\|,
$$
which also provides an upper bound for $\|t^{2(p+1)/d}u(t^{(p+1)/d}.,t)-AG_p\|$, if $t=L^n$. A similar argument to the linear case allows us to extend the result to all $t>1$.
\eop



\clearpage
\parskip 0pt
\baselineskip = 18pt


\end{document}